\documentclass[aps,prb,twocolumn,showpacs,preprintnumbers,amsmath,amssymb]{revtex4}
\usepackage{amssymb}
\usepackage{graphicx}
\usepackage{color}
\usepackage{dcolumn}
\usepackage{bm}

\begin{document}

\title{Evidence of Magnetically Driven Structural Phase Transition in Parent Compounds $R$FeAsO ($R$ = La, Sm, Gd, Tb): study of low-temperature X-ray diffraction}

\author{Yongkang Luo$^{1}$, Qian Tao$^{1}$, Yuke Li$^{1}$, Xiao Lin$^{1}$, Linjun Li$^{1}$, Guanghan Cao$^{1}$, Zhu-an Xu$^{1}$\footnote[1]{Electronic address: zhuan@zju.edu.cn},
Yun Xue$^{2}$, Hiroshi Kaneko$^2$, Andrey V. Savinkov$^2$,
Haruhiko Suzuki$^{2}$, Chen Fang$^{3}$, and Jiangping Hu$^{3}$}

\affiliation{$^{1}$Department of Physics, Zhejiang University, Hangzhou 310027, China\\
$^{2}$Department of Physics, Kanazawa University, Kakuma-machi, Kanazawa 920-1192, Japan\\
$^{3}$Department of Physics, Purdue University, West Lafayette,
Indiana 47907, USA}

\date{\today}

\begin{abstract}

We report measurements of structural phase transition of four
parent compounds $R$FeAsO ($R$ = La, Sm, Gd, and Tb) by means of
low-temperature X-ray diffraction (XRD). Magnetic transition
temperatures associated with Fe ions ($T_{N1}$) are also
determined from the temperature dependence of resistivity. As $R$
is changed from La, through Sm and Gd, to Tb, both the $c$-axis
and $a$-axis lattice constants decrease significantly. Meanwhile
both the structural phase transition temperature ($T_S$) and
$T_{N1}$ decrease monotonously. It is also found that the
temperature gap between $T_S$ and $T_{N1}$ becomes smaller when
the distance between FeAs layer becomes shorter. This result is
consistent with magnetically driven structural phase transition
and suggests that the dimensionality have an important effect on
the AFM ordering.

\end{abstract}


 \maketitle

\section{Introduction}

The recent discovery of superconductivity in layered
pnictide-oxide quaternary compounds $R$O$Tm$$Pn$ ($R$ =
lanthinides, $Tm$ = Fe, Ni, $Pn$ = P, As) has sparked enormous
interest in this class of
materials\cite{Hosono-LaP,Hosono-LaF,ChenXH-SmOF,WnagNL-CeOF,ZhaoZX-LnOD,WenHH-LaSr,WangC-GdTh,LuoJL-LaNi}.
Besides this 1111-type layered compounds, superconductivity was
subsequently discovered in other iron-based layered compounds with
similar FeAs(Se) layers, i.e., 122 systems\cite{Johrendt-BaK}, 111
systems\cite{LiFeAs}, and 11 systems\cite{WuMK-FeSe}. In all the
FeAs-based parent compounds, there is a  structural phase
transition in the temperature range 100-200 K, and a stripe-type
antiferromagnetic (AFM) ordering associated with Fe ions
accompanying the structural
transition\cite{WangNL-SDW,DaiPC-LaNeutron,BaoW-BaNeutron}.
Various chemical doping approaches can suppress the structural
transition and AFM order, and high-$T_c$ superconductivity
consequently appears. Meanwhile, low-$T_c$ superconductivity has
been observed in undoped FeP-based \cite{Hosono-LaP} and
NiAs(P)-based\cite{Hosono-NiAs,LuoJL-LaNi} compounds with similar
layered structure, but there is neither structural transition nor
AFM ordering associated with Fe(Ni) ions in these compounds. This
result implies that there is a relationship between structural
transition/AFM ordering and high-$T_c$ superconductivity.
Theoretically, the origin of the AFM order is still controversial.
There are two different physical pictures. One suggests that the
AFM order is a spin density wave (SDW) which is driven by Fermi
surface nesting between the electron pockets at M point and hole
pockets at $\Gamma$ point based on band structural
calculations\cite{WangNL-SDW,Mazin}. The other suggests that the
AFM order stems from the short range magnetic exchange coupling
between local moments\cite{Yildirim, SiQM, HuJP-Nematic,XiangT}.
However, regardless of the origin of the AFM ordering, the
theories suggest that the superconductivity is tied to the
magnetism in the FeAs-based
materials\cite{Mazin,Yildirim,XiangT,Seo2009}. The investigation
on the structural properties and AFM ordering of the parent
compounds can shed light on the mechanism of high-$T_c$
superconductivity.

The structural and magnetic transitions in the FeAs-based parent
compounds are deeply connected.  For the first discovered 1111
type systems, neutron scattering studies on the $R$FeAsO ($R$ =
La, Ce, Nd, and Sm) samples have found that the structural phase
transition occurs first  as temperature decreases, and then
magnetic ordering associated with Fe ions
follows\cite{DaiPC-LaNeutron,PhysC}, in contrast to the case in
122-type systems where both structural transition and AFM order
occur at the same temperature\cite{BaoW-BaNeutron}. Recent report
on isotope effect also shows positive isotope effect on both $T_c$
and AFM ordering temperature\cite{ChenXH-Isotope}. Some theoretic
studies proposed that the structural transition is directly driven
by the AFM order\cite{HuJP-Nematic,Yildirim,Xu}. In particular, a
theory based on a Heisenberg-type local moment exchange model
suggests that the structural transition can be driven by a nematic
Ising magnetic order due to the presence of intrinsic magnetic
frustration \cite{HuJP-Nematic}. The theory\cite{HuJP-Nematic}
predicts that the difference between the structural and magnetic
transition temperatures is controlled by the magnetic coupling
between layers: the difference becomes larger when the coupling is
weakened. Recent neutron experiment results in 4\% Ni-doped
BaFe$_2$As$_2$ have supported this prediction\cite{DaiPC-BaNi}.

In this paper, we report the investigation of structural phase
transition detected by means of low-temperature X-ray diffraction
(XRD) in the parent compounds LaFeAsO, SmFeAsO, GdFeAsO, and
TbFeAsO. The AFM order temperatures associated with Fe ions
($T_{N1}$) and the AFM ordering temperatures ($T_{N2}$) associated
with the magnetic rare earths Sm, Gd and Tb are also obtained by
measuring magnetic susceptibility and transport properties. A
systematic comparison of the structural transition temperature
($T_S$) with $T_{N1}$ is made. As $R$ is changed from La, through
Sm and Gd, to Tb, both the $c$-axis and $a$-axis lattice constants
decrease significantly. Meanwhile both $T_S$ and $T_{N1}$ decrease
monotonously. It is also found that the temperature gap
($T_S-T_{N1}$) becomes smaller when the distance between FeAs
layers becomes shorter. Therefore, our experimental results
provide concrete evidence supporting the theory proposed in
Ref.\cite{HuJP-Nematic} and suggest the dimensionality may have
important effect on the AFM ordering and on the superconductivity
mechanism as well.

\section{Experimental}

The polycrystalline $R$FeAsO ($R$ = La, Sm, Gd, Tb) samples were
synthesized by solid state reaction in vacuum using powders of
$R$As, $R_2$O$_3$ (for TbFeAsO, Tb$_4$O$_7$ was used instead),
FeAs and Fe$_2$As. $R$As was presynthesized by reacting
stoichiometric $R$ pieces and As powders in evacuated quartz tubes
at 1223 K for 24 hours. FeAs and Fe$_2$As were prepared by
reacting stoichiometric Fe powders and As powders at 1023 K for 20
hours. The powders of these intermediate materials were weighed
according to the stoichiometric ratio of $R$FeAsO respectively,
and then thoroughly mixed in an agate mortar. The mixtures were
pressed into pellets under a pressure of 4000 kg/cm$^2$. All the
processes were operated in a glove box filled with high-purity
argon. The pellets were sealed in evacuated quartz tubes and
heated uniformly at 1433-1453 K for 40 hours.

The sample purity was first checked by measurements of powder
X-ray diffraction (XRD) at room temperature using a D/Max-rA
diffractometer with Cu K$_{\alpha}$ radiation and a graphite
monochromator. Low temperature X-ray diffraction (LTXRD)
measurements for powder specimens were performed using the RINT
2500 system, Rigaku Co. An X-ray beam was generated by a rotating
Cu anode. The specimens were cooled by a $^4$He gas circulating
cryo-cooler and can be cooled down to about 10 K. The temperature
stability is better than 0.1\% during the LTXRD measurements. At
several temperatures entire profiles of reflection peaks were
measured with a step size of 0.01\textordmasculine and a
step-counting time of 6 s and refined by the Rietveld method using
the reported crystal structure. For some reflection planes X-ray
diffraction measurements with a step size of
0.005\textordmasculine and a step-counting time of 60 s were
performed to accumulate more counts at certain temperatures. From
the observed profile the
 $d$ value of (220) peak, the integrated intensity ($I. I.$) and also
the full-width-at-half-maximum ($FWHM$) were obtained. In these
analysis the profile was fitted to a Pseudo-Voigt function.

The electrical resistivity was measured by a standard
four-terminal method. The samples for transport property
measurements were cut into a thin bar. The temperature dependence
of d.c. magnetic susceptibility was measured on a Quantum Design
magnetic property measurement system (MPMS-5) under magnetic field
of 1000 Oe.

\section{Results and Discussion}

Fig. 1(a) shows the room-temperature XRD patterns of $R$FeAsO ($R$
= La, Sm, Gd, Tb) samples and Fig.1(b) shows the variations of
lattice constants $a$ and $c$ with the radius of $R$ ions. For all
the four parent compounds, the XRD peaks can be well indexed based
on a tetragonal cell with the space group of P4/$nmm$ (No. 129),
which indicates that the samples are in a uniform single phase
without obvious trace of impurity phases. As $R$ is changed from
La to Sm, Gd, and then Tb, all the peaks shift to larger
\textbf{2$\theta$'s} significantly, implying a remarkable
shrinkage of lattice in both $a$-axis and $c$-axis directions.
This result is consistent with the fact that the radius of $R$
ions decreases gradually as $R$ goes from the light to heavy rare
earth elements\cite{Shannon}. It can also be found from Fig.1(b)
that the $c$-axis shrinks slightly more quickly than the $a$ axis
does.

\begin{figure}
\includegraphics[width=0.95\columnwidth]{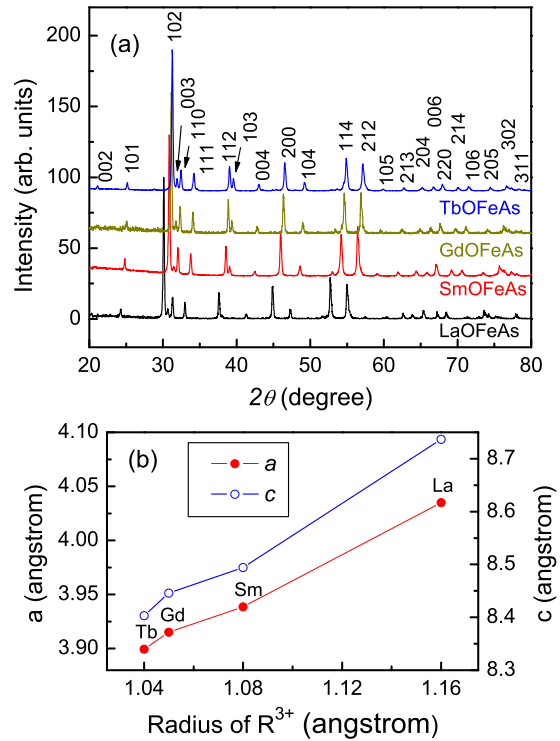}
\caption{(Color online) (a) X-ray powder diffraction pattern at
room temperature of $R$FeAsO ($R$ = La, Sm, Gd, Tb); (b) the
variations of lattice constants $a$ and $c$ with the radius of $R$
ions. The radius of $R$ ions is taken from Ref.\cite{Shannon}.}
\end{figure}

Fig. 2 shows the temperature dependence of resistivity for the
$R$FeAsO samples. The resistivity starts to drop around 120-150 K.
To show the drop position more precisely, we calculated the
derivative $d\rho/dT$ versus $T$ as shown in the inset of Fig.2.
The resistivity for LaFeAsO shows an upturn at low temperatures,
but it remains metallic for $R$FeAsO with $R$ = Sm, Gd, and Tb.
Similar resistivity has been observed in other 1111-type parent
compounds with magnetic rare earth elements\cite{McGuire}. Such a
difference in low temperature resistivity has not been well
understood yet. We define the characterization temperature
$T_{N1}$ as the peak position in the curves of $d\rho/dT$ versus
$T$. As shown in the inset, $T_{N1}$ decreases significantly as
$R$ is changed from La to Sm, Gd, and Tb. Neutron studies have
confirmed that the resistivity anomaly is caused by the structural
phase transition and the following formation of antiferromagnetic
SDW state\cite{DaiPC-LaNeutron}. Previous reports have proposed
that the peak position in $d\rho/dT$ corresponds to the AFM
ordering of Fe ions moments rather than the structural phase
transition\cite{McGuire,Klauss}. Indeed, the studies of neutron
diffraction reported that the AFM ordering temperature of LaFeAsO
is about 135 K, about 20 K lower than the structural phase
transition temperature $T_S$ of 158 K\cite{DaiPC-LaNeutron}. The
$T_{N1}$ value of LaOFeAs determined from the resistivity is 132
K, consistent with the  AFM order temperature reported by the
neutron diffraction. The AFM order temperature determined from the
measurements of M\"{o}ssbauer spectroscopy and $\mu$SR relaxation
is also in agreement with $T_{N1}$, the peak temperature in the
$d\rho/dT$ versus $T$ curves\cite{Klauss}. Thus we can regard
$T_{N1}$ as the characteristic temperature at which the magnetic
moments of Fe ions become AFM ordered. For $R$ = Sm, Gd, and Tb,
$T_{N1}$ is 133 K, 128 K, and 122 K, respectively.

\begin{figure}
\includegraphics[width=0.9\columnwidth]{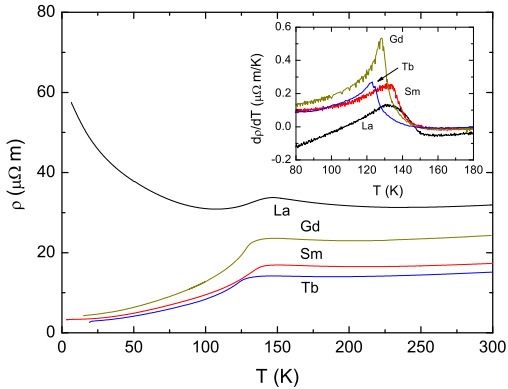}
\caption{(Color online) Temperature dependent resistivity of
$R$FeAsO ($R$ = La, Sm, Gd, Tb). Inset: the derivative of
resistivity($d\rho/dT$) as a function of temperature.}
\end{figure}

\begin{figure}
\includegraphics[width=0.9\columnwidth]{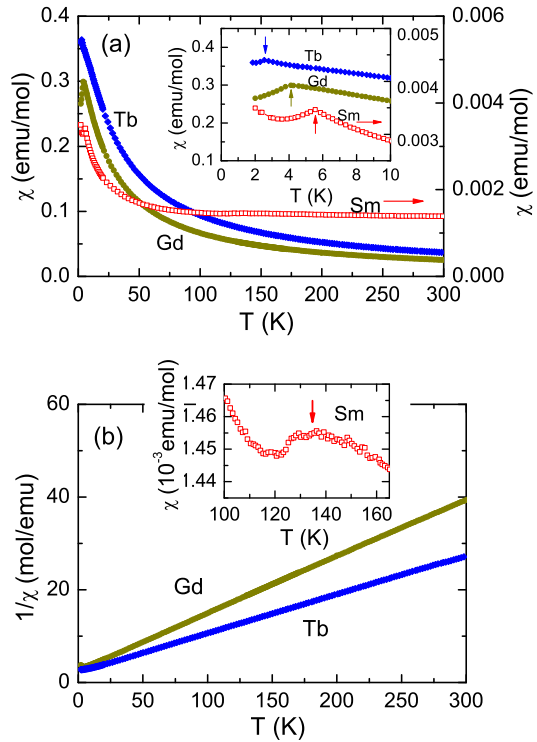}
\caption{(Color online) (a) Temperature dependence of magnetic
susceptibility measured under $H$ of 1000 Oe for SmFeAsO, GdFeAsO
and TbFeAsO. Inset: enlarged plot for low temperatures to show the
AFM transitions of $R^{3+}$ ions; (b) The plot of $\chi^{-1}$
versus $T$ for GdFeAsO and TbFeAsO. The linear behavior for $T$
$>$ $T_{N2}$ means that $\chi$ can be fitted by the Curie-Weiss
law very well. Inset: Enlarged plot for SmFeAsO to show the kink
around the AFM ordering temperature ($T_{N1}$) of Fe ions.}
\end{figure}

In order to obtain more information about the magnetism associated
with Fe ions and $R^{3+}$ ions as $R$ is magnetic rare earth
elements other than La, the magnetic susceptibility of $R$FeAsO
($R$ = Sm, Gd, Tb) was measured under the magnetic field of 1000
Oe, as shown in Fig.3. From the temperature dependence of magnetic
susceptibility shown in Fig. 3(a) and its inset, it can been found
that the GdFeAsO and TbFeAsO have much larger magnetic
susceptibility compared to SmFeAsO because Gd$^{3+}$ and Tb$^{3+}$
ions have much larger magnetic moments. At low temperatures, clear
phase transitions caused by the AFM ordering of the magnetic
moments of $R$ ions can be found at $T_{N2}$. The AFM order
temperature of $R$ ions, i.e., $T_{N2}$ determined in our
measurements, is 5.56 K, 4.11 K, and 2.54 K for SmFeAsO, GdFeAsO,
and TbFeAsO, respectively, consistent with previous
reports\cite{WangC-GdTh,SmAFM,TbAFM}. For SmFeAsO, as shown in the
inset of Fig.3(b), a kink associated with the AFM ordering of Fe
ions can be observed around 134 K, consistent with the $T_{N1}$
value of 133 K within the experimental error. For GdFeAsO and
TbFeAsO, this kind of AFM order associated with Fe ions is buried
in the large magnetic signals from the $R^{3+}$ ions. The magnetic
contributions from the $R^{3+}$ ions obey the Curie-Weiss law very
well. As shown in the Fig.3(b), the inverse of $\chi$ increases
strictly linearly with $T$ as $T > T_{N2}$. For the SmFeAsO
sample, the magnetic susceptibility does not exhibit the
Cuire-Weiss behavior because the contribution from the Fe ions
which is linearly dependent on temperature is comparable to the
contribution from the Sm$^{3+}$ ions\cite{GMZhang}. By fitting the
Curie-Weiss law, we obtained that the effective magnetic moments
$p_{eff}$ are 8.0, and 9.7 $\mu_B/f.u.$ for GdFeAsO and TbFeAsO
respectively, which are consistent with the theoretical values of
magnetic moments of free Gd$^{3+}$ and Tb$^{3+}$ ions. If we
subtract the Curie-Weiss term which should originate from the
contributions of $R^{3+}$ ions, we can also find a slight drop in
the subtracted term ($\chi-\chi_{CW}$) around $T$ of 120-140 K for
GdFeAsO and TbFeAsO (not shown here), where $\chi_{CW}$ is the
Cuire-Weiss fitting function. But it is very hard to distinguish
whether such a drop in $\chi-\chi_{CW}$ occurs at the AFM order
temperature of Fe ions or the structural phase transition
temperature. Therefore, we will take $T_{N1}$ as the transition
temperature associated with the AFM order of Fe ions.

As mentioned above, the neutron studies\cite{DaiPC-LaNeutron} have
revealed that a structural phase transition occurs just before the
AFM ordering in LaFeAsO. It is generally believed that the
structure of the parent compounds $R$FeAsO transforms from
tetragonal to orthorhombic when the temperature is lower than the
structural transition temperature $T_S$. Such a structure phase
transition can be detected by the splitting of (220) peak in the
low temperature x-ray diffraction. Fig. 4 shows the temperature
dependence of the (220) peak $d$ value for $R$FeAsO samples. The
inset shows the intensity of (220) peak before and after the
structural phase transition. It can be seen that the (220) peak
splits into two peaks when the temperature is lower than the
structural phase transition temperature. From the temperature
dependence of (220) peak $d$-value, the structural phase
transition temperature $T_S$ can be easily determined as the
splitting point of this peak. As $R$ changes from La to Tb, the
(220) peak $d$ value decreases significantly, and $T_S$ decreases
as well.

\begin{figure}
\includegraphics[width=1.0\columnwidth]{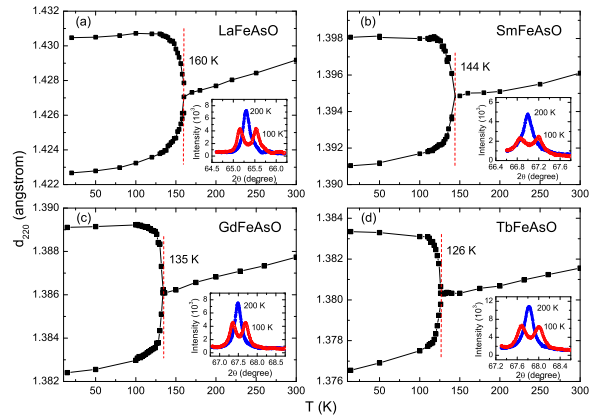}
\caption{(Color online) The temperature dependence of the (220) peak
$d$ value for LaFeAsO (a), SmFeAsO (b), GdFeAsO (c) and TbFeAsO (d).
The insets show the intensity of (220) peak before (blue lines) and
after (red lines) structure phase transition. }
\end{figure}

We summarize the variations of the structural phase transition
temperature ($T_S$) and AFM ordering temperature ($T_{N1}$) in
Fig. 5. The structural and physical parameters for these four
parent compounds are also listed in Table I. As $R$ is changed
from La, through Sm and Gd, to Tb, not only $T_S$ and $T_{N1}$
decrease significantly, but the temperature gap between $T_S$ and
$T_{N1}$ also becomes smaller, i.e., the AFM transition occurs at
the temperature closer to the structural phase transition
temperature. Actually the structural phase transition and AFM
ordering happen simultaneously in the more three-dimensional 122
parent compounds like BaFe$_2$As$_2$.

\begin{figure}
\includegraphics[width=0.9\columnwidth]{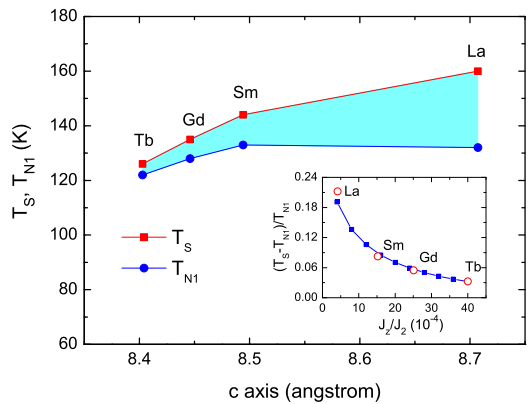}
\caption{(Color online) Plot of structural phase transition
temperature $T_S$, and magnetic transition temperature $T_{N1}$
associated to the AFM order of Fe ions, versus the $c$-axis for
$R$FeAsO ($R$ = La, Sm, Gd, Tb). Inset: plot of
$(T_S-T_{N1})/T_{N1}$ versus $J_z/J_2$. The open circles denote
the experimental data of $(T_S-T_{N1})/T_{N1}$, and the solid
squares denote the theoretical values. See text for details.}
\end{figure}

The experimental results can be understood within the theory
proposed in Ref.\cite{HuJP-Nematic} where an effective
Heisenberg-type magnetic exchange model, the so called
$J_1$-$J_2$-$J_z$ model with $J_1$, $J_2$  and $J_z$ being the
in-plane nearest neighbor(NN), in-plane next nearest neighbor(NNN)
and out-of-plane magnetic exchange couplings respectively,
explains both the structural and magnetic transitions associated
with the FeAs layers. In this model, a collinear AFM ground state
is obtained when $J_1$ is less than 2$J_2$, and a nematic Ising
magnetic order transition that breaks lattice rotational symmetry
takes place at a temperature equal to or higher than the collinear
AFM transition temperature\cite{HuJP-Nematic}. By including the
lattice and spin coupling, the nematic order naturally produces an
orthorhombic lattice distortion. Therefore, the model captures
both structural and magnetic transitions and suggests the
structural transition be driven magnetically. A quantitative
prediction of this model is that the difference between the
structural and AFM transition temperature is determined by the
ratio between $J_z$ and $J_2$, i.e., $J_z/J_2$. The AFM transition
temperature, $T_{N1}\propto J_2 /\log(J_2/J_z)$. Using the
calculated results in Ref.\cite{HuJP-Nematic} and assuming that
$J_1 \approx J_2$, we find that the ratio $J_z/J_2$ for the four
typical parent compounds (La, Sm, Gd, Tb)FeAsO are (4.10, 15.2,
25.1, 40.0)$\times 10^{-4}$. The out-of-plane magnetic exchange
coupling increases quickly as the lattice constant along $c$-axis
deceases, and therefore the temperature difference between $T_S$
and $T_{N1}$ should decrease according to this model. The result
is plotted in the inset of Fig.5. The experimental data of
$(T_S-T_{N1})/T_{N1}$ are in agreement with the theoretical
calculations. By measuring the spin wave gap around the wavevector
(0, $\pi$, 0) as shown in Ref.\cite{zhaojun} in future neutron
scattering experiments, the value of $J_z$ can also be
independently obtained.

\begin{table}
\caption{\label{tab:table1}Structural and Physical parameters of
$R$FeAsO ($R$ = La, Sm, Gd, and Tb)}
\begin{center}
\begin{ruledtabular}
\begin{tabular}{cccccccc}
       Sample  & $r$($R^{3+}$) (\AA) & $a$ (\AA) & $c$ (\AA) & $T_{S}$ (K) & $T_{N1}$ (K) &  $T_{N2}$ (K) \\ \hline
       LaFeAsO &                 1.16  &   4.0349  &  8.7366   &     160     &      132     &    -      \\
       SmFeAsO &                 1.08  &   3.9385  &  8.4941   &     144     &      133     &   5.56    \\
       GdFeAsO &                 1.05  &   3.9151  &  8.4660   &     135     &     128    &   4.11    \\
       TbFeAsO &                 1.04  &   3.8994  &  8.4029   &     126     &      122     &   2.54
\end{tabular}
\end{ruledtabular}
\end{center}
\end{table}

\section{Conclusion}

In summary, we have studied the structural phase transition of
four parent compounds $R$FeAsO ($R$=La, Sm, Gd, Tb) by measuring
low-temperature X-ray diffractions. As $R$ is changed from La to
Tb, $T_s$ and $T_{an}$ as well as the $c$ axis decrease
significantly. Furthermore, the temperature difference between
$T_S$ and $T_{N1}$ becomes smaller as the $c$ axis becomes
shorter. According to the theoretical calculations proposed in
Ref.\cite{HuJP-Nematic}, the out-of-plane magnetic exchange
coupling increases quickly with the decreasing $c$-axis lattice
constant and therefore the temperature difference between $T_S$
and $T_{N1}$ is significantly influenced. The experimental data of
$(T_S-T_{N1})/T_{N1}$ are in agreement with the theoretical
calculations. This result supports the theoretical proposal that
the structural phase transition is driven by a nematic Ising
magnetic order due to the presence of intrinsic magnetic
frustration, and indicates that the dimensionality could have
important effect on the AFM ordering, and the magnetism may play
an important role in high-$T_c$ superconductivity of the iron
pnictides.

\section*{Acknowledgments}

This work is supported by the National Science Foundation of China
(No. 10634030), PCSIRT (No. IRT0754), and the National Basic
Research Program of China (No.2006CB601003 and 2007CB925001).


\end{document}